\documentclass{article}
\usepackage{amsmath,amssymb}
\usepackage{graphicx}
\usepackage{dcolumn}
\usepackage{bm}
\usepackage{hyperref}
\usepackage{tikz}
\usepackage{tikz-cd}

\begin{document}

\title{Free to Interacting Map for Crystalline SPT Phases:
Equivariance vs Crystalline Equivalence}

\author{Daniel Sheinbaum\\
\small Division of Applied Physics, CICESE 22860, Ensenada, BC, Mexico\\
\small\texttt{daniels@cicese.mx}
\and
Omar Antol\'in Camarena\\
\small Instituto de Matem\'aticas, National Autonomous University of Mexico,\\
\small 04510 Mexico City, Mexico\\
\small\texttt{omar@im.unam.mx}}

\date{\today}

\maketitle

\begin{abstract}
Freed and Hopkins developed an ansatz for classifying interacting SPT phases using invertible field theories \cite{Freed-Hopkins-SPT} with a natural Free to Interacting (FTI) map from free fermion phases. This ansatz has been generalized to include crystalline phases and a crystalline equivalence principle (CEP) \cite{Freed-Hopkins-Spatial}. However, motivated by failure of the CEP for weak free fermions \cite{SA-CEP} and the FTI, here we generalize the original Freed and Hopkins ansatz to a fully equivariant version for symmorphic crystallographic symmetries and show there is a natural equivariant FTI map from symmorphic crystalline weak free fermions. We further discuss why this equivariant ansatz is both mathematically and physically more natural than the one in \cite{Freed-Hopkins-Spatial} and why full equivariance should hold over the CEP.
\end{abstract}


\section{Introduction}\label{sec:Introduction}
Topological phases of matter have been classified using generalized cohomology theories like $K$-theory for free fermions \cite{Kitaev} including both internal and spatial symmetries \cite{Freed-Moore}. When including short range interactions the notion of symmetry protected topological (SPT) phase with internal symmetry group $H$ was introduced and classified first in \cite{Wen-SPT}, using the group cohomology of $H$. However the community considered that this classification was incomplete and we now have various attempts at a generalization \cite{Kapustin-SPT}, \cite{Freed-Hopkins-SPT}, \cite{Gaiotto-SPT}, \cite{Xiong-SPT}, \cite{Kubota-SPT} among others, all under the paradigm of what is known as Kitaev's conjecture. In particular the Ansatz of Freed and Hopkins \cite{Freed-Hopkins-SPT} classifies interacting SPTS employing invertible field theories. One advantage of this approach is that it has a natural Free to Interacting (FTI) map, which has been extended to weak phases \cite{ADKPSS-FTI}. Crystalline SPT phases arise when the symmetry protecting the phase is a crystallographic symmetry (discrete translations, rotations and reflections). In this work we will only consider \textit{symmorphic} crystalline symmetries for simplicity. For free fermions, one can derive a classification from first principles which yields equivariant $K$-theory \cite{Freed-Moore}. In their pioneering work \cite{Thorngren-Else}, Thorngren and Else developed a procedure to gauge spatial symmetries, leading to a crystalline equivalence principle (CEP), which states you can treat spatial symmetries as if they were internal and thus crystalline SPT's are classified using group cohomology of the crystallographic symmetry group. Freed and Hopkins further proposed a generalization of their ansatz to include spatial symmetries and such that it has a CEP \cite{Freed-Hopkins-Spatial}. Nevertheless, the authors of the current manuscript showed that the CEP does not hold in the case of crystalline weak free fermion phases \cite{Crystalline-FF}. We first motivate, using a crystalline version of Kitaev's conjecture why a different extension of the ansatz should be preferred. We next elaborate why the ansatz in \cite{Freed-Hopkins-Spatial} has a natural FTI map to the wrong theory for free fermions. We further proceed to give such a genuinely equivariant extension of the original Freed and Hopkins's ansatz (in the symmorphic case), which does not satisfy a CEP, and show how there is a natural equivariant FTI map to crystalline free fermions. We note from the outset that our arguments do not prove conclusively that the CEP should fail in the interacting case, instead they point towards the naturality of both full crystalline equivariance and an equivariant FTI map for classifying SPTs.

\section{The CEP and its failure for weak free fermions}\label{sec:CEP}
Let us consider $(d+1)$-dimensional system with crystalline symmetry group $G$ as well as internal symmetry group $H$. The CEP \cite{Thorngren-Else} coupled to the group cohomology classification \cite{Wen-SPT} leads to the following classification of $(d+1)$-dimensional $H\times G$ Bosonic crystalline SPT phases 
\begin{align}\label{eq:TE}
    H^{d+2}(B(H\times G); \mathbb{Z}),
\end{align}
where $B(H\times G)$ is the classifying space of the group $H\times G$. Note that $G$ and $H$ are treated on equal footing. Furthermore, the different actions of $G$ and $H$ on the space of Hamiltonians (how the symmetries are represented in the Hilbert space) are in a sense reduced to their homotopy bare essentials, which take the form of the classifying space $B(H\times G)$. The erasure of the difference in the actions is, in broad terms, what gives rise to the CEP. However, this erasure of the distinct actions simply does not occur for free fermions \cite{Crystalline-FF}, on the one hand because equivariant $K$-theory is not of Borel type for neither internal nor spatial symmetries with being Borel the way the CEP usually manifests itself. On the other, it is not only so that neither is Borel but they are not equivalent in general since the internal symmetries mainly shift the degree of the group (and make the vector bundles either real or complex) and the point group $P$ acts on the real space unit cell torus $\mathbf{T}^d$, which $H$ does not. Principles must hold in every situation and thus we are led to ask, if it fails for free fermions why would it hold when including short-range interactions?  

\section{Crystalline Kitaev's conjecture and to Borel or not Borel}
At this point Kitaev's conjecture \cite{Kitaev-SPT} enters our story. Kitaev's conjecture states that the set of $(d+1)$-dimensional gapped short-range entangled Hamiltonians $\mathcal{S}^{d}(H)$ with internal symmetry group $H$ form a spectrum \cite{Hatcher-Alg-Top}, i.e. they yield a cohomology theory such that SPT phases, which are the path-components of $\mathcal{S}^{d}(H(s))$ satisfy
 \begin{equation}
     \pi_0(\mathcal{S}^{d}(H))= D(H)^d(\cdot).
\end{equation}
 For some generalized cohomology theory $D(H)$. The natural generalization of this conjecture to include discrete translation symmetries would say \cite{ADKPSS-FTI}
  \begin{equation}
     \pi_0(\mathcal{S}^{d}(H,\mathbb{Z}^d))= D(H)^d(\mathbf{T}^d),
\end{equation}
with $\mathbf{T}^d$ being the real space unit cell torus. Free fermions are a fantastic testbed for the validity and possible extensions of conjectures such as Kitaev's and the CEP. In particular if we were to formulate a version which included point group crystalline symmetries, for it to be satisfied by free fermions, the spectrum $D(H)$ must be made equivariant with respect to $P$-action on $\mathbf{T}^d$. This implies that the crystalline version of Kitaev's conjecture for a crystalline symmetry group $G$ is

 \begin{equation}
     \pi_0(\mathcal{S}^{d}(H,G))= D(H)_P^d(\mathbf{T}^d).
\end{equation}
We get some confirmation from the fact that it holds for the case of crystalline Free fermions, where you have $D(H(s))_P^d = \mathit{KO}_P^{d+s-2}$ ---a priori, this may not seem to match Kitaev's conjecture since the degree is $d+s-2$ rather than $d$, but this only indicates that the free fermion phases are classified by the spectrum $\Sigma^{s-2} \mathit{KO}_P$. Note that we have restored the $s$ in the notation to show how $H$ enters in $K$-theory. A mathematically natural question which arises for equivariant cohomology theories is if they are of Borel type, that is, if for a general $P$ space $X$, we have 
\begin{equation}
     D(H)^d_P(X) = D(H)^d(X\times_P EP),
\end{equation}
or in its simplest version, when $X$ is a point if
\begin{equation}
     D(H)^d_P(pt) = D(H)^d(BP),
\end{equation}
where $BP$ is the classifying space of $P$, arising here as the Borel construction of the trivial $P$-action on a point. To be Borel is in a sense to forget some details of the symmetry i.e. the exact group action and instead to take the action up to homotopy \cite{Greenless-Equivariant}.  We note that most equivariant cohomology theories are not of Borel type, i.e it is highly unlikely that if we choose one at random it turns out to be Borel. The original choice of spectrum for $D^d(H)$, the group-cohomology $K(H,d+2,\mathbb{Z)}$ spectrum is naturally extended to an equivariant version, which satisfies our crystalline Kitaev's conjecture and is also naturally extended to a Borel theory i.e. 

\begin{align}
     D(H)^d_P(\mathbf{T}^d) &= H^{d+2}_{P\times H}(\mathbf{T}^d;\mathbb{Z}),\\
     &= H_{H}^{d+2}(\mathbf{T}^d\times_P EP;\mathbb{Z}),\\
     &= H^{d+2}(B(G\times H);\mathbb{Z})
\end{align}
and hence carries a CEP as elucidated in \cite{Thorngren-Else}.

On the other hand, the only classification which has been derived rigorously from first principles is that of free fermions and $P$-equivariant $K$-theory of $\mathbf{T}^d$, which is not Borel with respect to either internal nor crystalline symmetries. Note that nevertheless it also satisfies our crystalline Kitaev's conjecture. What about other choices $D(H)$, is their natural $P$-equivariant extension also Borel? For the choice of the Freed-Hopkins Ansatz $\mho_{H(s)}^{d+2}$ of invertible TQFTs with tangential structure given by $H(s)$ \cite{Freed-Hopkins-SPT} this is not so, as we shall elucidate below, when we construct the natural $P$-equivariant version. Thus the paradigm of a crystalline Kitaev's conjecture combined with the Freed-Hopkins ansatz naturally leads to a non-Borel $P$-equivariant generalized cohomology theory which does not possess a CEP.\\
There is by now an extension of the Freed-Hopkins ansatz to include spatial symmetries \cite{Freed-Hopkins-Spatial}, \cite{Arun-SPT}, \cite{Yu-GCEP}. We show below there are naturality issues with this ansatz when combined with what ought to be the FTI map.

\section{The wrong FTI map}
Physically, since the set of free-fermion Hamiltonians is a special case of Interacting fermion Hamiltonians with zero interaction, any theory which attempts to classify interacting fermionic SPT phases, should be equipped with a FTI map from the $K$-theoretic free fermion phases, predicting what happens to free fermion phases once you allow adiabatic evolution through systems with interactions, but without closing the gap.

In what follows we use the framework of \cite{ADKPSS-FTI}, where the internal symmetries are represented by a Lie group $H(s)$ that combines both the Euclidean symmetry group $O(d+1)$ (to account for Wick rotation) and CRT symmetries. Although not derived from first principles, the existence of an FTI map \cite{Freed-Hopkins-SPT}, \cite{ADKPSS-FTI}, \cite{DKS-Bott}
\begin{align}\label{eq:F2I}
FTI_{\mathit{weak}}:KO^{d{+}s-2}(\mathbf T^d) \longrightarrow \mho_{H(s)}^{d+2}(\mathbf T^d)
\end{align}
is a strong motivation in favor of the Freed-Hopkins ansatz as it is precisely what allows the Atiyah-Bott-Shapiro map to connect $K$-theory and its real variants to the different types of invertible TQFTs with tangential structure given by $H(s)$, that is, to $\mho_{H(s)}^{d+2}(\mathbf T^d)$. Note that this is the only extension beyond group cohomology which, to our knowledge, has a FTI map. Also note that we have formulated free fermion phases using the real space torus through $T$-duality \cite{ADKPSS-FTI}. 

We can now ask: Does something similar happen for spatial symmetries? i.e. is there a natural FTI map from the crystalline free fermion phases $KO_P^{d{+}s-2}(\mathbf T^d)$ \cite{Thiang-Crystallographic-1} to a $P$-equivariant generalization of $\mho_{H(s)}^{d+2}(\mathbf T^d)$? Let us first consider the extension by Freed and Hopkins of their own work \cite{Freed-Hopkins-SPT} for spatial symmetries \cite{Freed-Hopkins-Spatial}. They propose that crystalline SPT phases are given by
\begin{equation}
   E^{hP}(T^d):=[S^0, \Sigma^2 I_\mathbb{Z}(MTH(s))\wedge T^d_+]^{hP},
\end{equation}
where the superscript $hP$ indicates taking homotopy orbits for the $P$-action, and $I_\mathbb{Z}$ denotes Anderson duality. Note that this formula is covariant as functor of the torus, so it represents a type of homology rather than cohomology ---it is Borel-Moore homology, in fact.

Currently there is only one known strategy for constructing a natural FTI map \cite{Freed-Hopkins-SPT}: \cite{ADKPSS-FTI} using the twisted Atiyah-Bott-Shapiro map,
\begin{equation}
    \phi: MTH(s)\longrightarrow \Sigma^{-s}KO,
\end{equation}
which sends the Clifford structure of the tangent bundle of the manifold to a twisted Clifford Dirac operator. Applying Anderson duality, which reverses the direction of the map,  we obtain a candidate FTI map, which we will argue is not what we want: \begin{equation}
    FTI_{\mathit{wrong}}:\,[S^0,\Sigma^2 I_\mathbb{Z}(\Sigma^{-s}KO)\wedge T^d_+]^{hP}\longrightarrow E^{hP}(T^d).
\end{equation}
The problem with this map is that the domain of $FTI_\mathit{wrong}$ is not what classifies crystalline free fermions! In fact it is a sort of twisted Borel equivariant version of the correct crystalline free fermion group $KO_P^{d+s-2}(\mathbf T^d)$, where the twist is given by the $P$-representation $V$. Indeed, Anderson self-duality of $KO$ tells us that $I_\mathbb{Z}(KO) \simeq \Sigma^{-4} KO$. Also, we can use Atiyah duality to move the torus to the left; because the tangent bundle of the torus is trivial, Atiyah duality says that the dual of $T^d_+$ is $S^{-V} \wedge T^d_+$, so:
\begin{equation}
    [S^0,\Sigma^2 I_\mathbb{Z}(\Sigma^{-s}KO)\wedge T^d]^{hP} \cong [S^V \wedge T^d_+, \Sigma^{s-2}KO]^{hP}
\end{equation}
Next, since $KO$ has a trivial action of the point group $P$, the homotopy fixed points becomes a Borel construction:
\begin{equation}
    [S^{-V} \wedge T^d_+, \Sigma^{s-2}KO]^{hP} \cong [(S^{-V} \wedge T^d_+ \wedge EP_+)/P, \Sigma^{-2}KO].
\end{equation}
Notice that if instead of the representation sphere $S^{-V}$, one had the trivial $P$-action on $S^{-d}$, one could lift it outside of the quotient by $P$ and this would simplify to $KO^{d+s-2}((T^d \times EP)/P) \cong KO^{d+s-2}(BG)$, which is Borel $K$-theory of the torus. The representation sphere makes this a twisted version of that $K$-theory group.

Given that the ansatz in \cite{Freed-Hopkins-Spatial} has naturally a would-be FTI map to the wrong group, one that is Borel with respect to spatial symmetries for free fermions, it strongly suggests, though not definitively, that to have a natural FTI map to the non-Borel 
$KO_P^{d{+}s-2}(\mathbf T^d)$ we require a non-Borel extension of \cite{Freed-Hopkins-SPT}. Nevertheless, we have not shown conclusively that such a map from the ansatz in \cite{Freed-Hopkins-Spatial} does not exist.\\
From a physical point of view the wrong FTI map opens up a bizarre possibility, for there could be elements in $E^{hP}(T^d)$ in the image of $FTI_{\mathit{wrong}}$ on a ``spurious" free phase i.e. broadly understood as an element in the domain of $FTI_\mathit{wrong}$ which is not in $KO_P^{d+s-2}(\mathbf T^d)$. If such a phase in $E^{hP}(T^d)$ were to exist, it would mean that there is an interacting phase that wants to be like a non-existent free phase. However, once again to conclude this would require explicit computations of both groups to search for such spurious phases, which we leave for future work.

\section{A non-Borel Freed-Hopkins Ansatz}
As we mentioned previously Freed and Hopkins's Ansatz for a spectrum  $\mho_{H(s)}^{d+2}$ classifies deformation classes of extended, reflection positive invertible field theories \cite{Freed-Hopkins-SPT}. Skipping to the punchline, the natural extension of this spectrum would be one which classifies deformation classes of extended, reflection positive invertible field theories with a $P$-action. For this we need to do a small digression and disassemble the spectrum $\mho_{H(s)}^{d+2}$ itself; It consists of two pieces: The first is the spectrum $MTH(s)$ made out of extended invertible field theories with tangential structure $H(s)$ and the second is the Anderson dual $I_{\mathbb{Z}}$, which imposes reflection positivity (the Wick rotated version of unitarity). Viewed in terms of homotopy theory we have, for a nice, abstract topological space $X$, the equivalence
\begin{equation}
    \mho_{H(s)}^{d+2}(X) = [X\wedge MTH(s),\Sigma^{d+2}I_{\mathbb{Z}}].
\end{equation}
To make our desired generalization we simply need to modify the first piece, $MTH(s)$ into a spectrum that considers the $P$-action on manifolds. Fortunately, this was recently done for finite $P$ by Galatius and Szücs \cite{Galactus-Equivariant} and for different tangential structures. Note that the extension of Galatius and Szücs is not Borel with respect to the $P$-action and luckily the point group $P$ is indeed finite. Let us denote this spectrum $MTH(s)_P$.  This is of course not the most general case since for example reflections can square to fermion parity $(-1)^F$. To fix this would require a de-Borelianization of the work by Debray \cite{Arun-SPT} and modifying \cite{Galactus-Equivariant} to include the mixing of internal and spatial symmetries, hence we leave it for future work. For non-symmorphic crystallographic SPTs this would involve some form of twisting this spectrum and leave it for future work as well. Nevertheless this spectrum is sufficient to construct an equivariant FTI map for all symmorphic examples of \cite{Freed-Moore}, as we show below. We thus define, for a nice $P$-CW complex $X$, the truly equivariant Freed-Hopkins Ansatz as

\begin{equation}
    \mho_{H(s),P}^{d+2}(X) = [X\wedge MTH(s)_P,\Sigma^{d+2}I_{\mathbb{Z}}]_P.
\end{equation}

Note that the $P$-action on $\Sigma^{d+2}I_{\mathbb{Z}}$ is trivial so that we are actually looking at $P$-invariant maps. In particular this would mean that the genuinely equivariant Freed–Hopkins ansatz assigns as the set of $(d+1)$-dimensional crystalline SPT phases with internal group $H(s)$ and crystalline symmetry group $G$ the following
 \begin{equation}\label{eq:Equiv-Ansatz}
     \pi_0(\mathcal{S}^{d}(H(s),G))= \mho_{H(s),P}^{d+2}(\mathbf{T}^d).
\end{equation}
We emphasize that this new proposal for the classification of crystalline SPT phases does not have a CEP, because whereas it is Borel with respect to the internal symmetry group $H(s)$ it is not Borel with respect to $P$ (nor $G$) as the Borel construction of either does not appear in it. 

\section{Equivariant FTI map}

To assemble an equivariant FTI note that there is an equivariant ABS map \cite{Joachim-Equivariant-ABS}
\begin{align}
\phi_P:MSpin_P^{c} \longrightarrow K_P.
\end{align}
So that the twisted ABS map constructed in \cite{Freed-Hopkins-SPT}, subsection 9.22 extends to the equivariant case, yielding a natural equivariant map between $P$-spectra.
\begin{align}\label{eq:Twisted-Equiv-ABS}
\phi_P:MTH(s)_P \longrightarrow \Sigma^{s}KO_P.
\end{align}

The other necessary ingredient is that $P$-equivariant $K$-theory is Anderson self-dual \cite{Joachim-Equivariant-Anderson} i.e.
\begin{equation}\label{eq:Equiv-Anderson}
    I_{\mathbb{Z}}(KO_P) \simeq \Sigma^4 KO_P.
\end{equation}

 Thus we can readily repeat step by step the construction in \cite{ADKPSS-FTI} in the equivariant case using crystalline $T$-duality \cite{Thiang-Crystallographic-1}, \cite{Thiang-Crystallographic-2}, the map \ref{eq:Twisted-Equiv-ABS} and eq (\ref{eq:Equiv-Anderson}) yielding 

\begin{align}\label{eq;Equiv-F2I}
FTI_{P}:KO_P^{d{+}s-2}(\mathbf T^d) \longrightarrow \mho_{H(s),P}^{d+2}(\mathbf T^d).
\end{align}

We claim that (\ref{eq;Equiv-F2I}) is natural, arising from the obvious generalization of the ABS map to the $P$-equivariant and $H(s)$-twisted case, giving favor to our new proposal $\mho_{H(s),P}^{d+2}(\mathbf T^d)$ over that of \cite{Freed-Hopkins-Spatial}. As in \cite{Freed-Hopkins-SPT},\cite{ADKPSS-FTI}, the kernel of this map tells us which symmorphic crystalline free fermion phases are killed by interactions and the cokernel tells us which interacting symmorphic crystalline SPT phases in $\mho_{H(s),P}^{d+2}(\mathbf T^d)$ are interaction-enabled crystalline phases.

\section{Discussion}
We have argued, starting from a crystalline Kitaev's conjecture, that cohomology theories with a genuine equivariant generalization should be preferred over their Borel counterparts to fit crystalline free fermions into the paradigm, we then showed that the ansatz in \cite{Freed-Hopkins-Spatial} has a natural FTI map to the wrong free fermion group and this opens up the possibility of unphysical spurious free phases stable to interactions appearing in the classification of \cite{Freed-Hopkins-Spatial}, though we have not provided an example to such a fact. We then constructed a crystalline equivariant version of the Freed-Hopkins Ansatz $\mho_{H(s),P}^{d+2}(\mathbf{T}^d)$ (\ref{eq:Equiv-Ansatz}), which is a natural equivariant generalization of \cite{Freed-Hopkins-SPT} but, like the case of free fermions, lacks a CEP. We also show this theory has a natural equivariant Free to Interacting map (\ref{eq;Equiv-F2I}).\\
What about explicit computations of the groups representing SPT phases in our new proposal? This is in a sense the biggest drawback of our approach which is that, though not too difficult to define, computations, even in the most simple cases, seem quite formidable. Nevertheless, we believe that we have provided strong arguments, both on the mathematical and on the physical side, to favor our proposal despite its potential computational difficulty. This proposal states that full equivariance for crystalline symmetries should be taken as a principle at the cost of the CEP not being among these principles. Viewing the glass half full, we hope this is sufficient motivation for the algebraic topology community to further develop methods for computing these groups, so that we can make explicit comparisons to the ansatz of \cite{Freed-Hopkins-Spatial}.
\providecommand{\noopsort}[1]{}\providecommand{\singleletter}[1]{#1}%

\end{document}